\documentclass{sig-alternate}
\usepackage{fdiaz}
\usepackage{array,multirow,graphicx}
\usepackage{clrscode}
\usepackage{subcaption}
\usepackage[hyphens]{url}

\usepackage[usenames,dvipsnames]{xcolor}
\makeatletter
\def\@copyrightspace{\relax}
\makeatother

\newcommand{\bhat}[2]{\mathcal{B}(#1,#2)}

\newcommand{\corpuslm}[0]{\theta_C}
\newcommand{\depth}[0]{d}
\newcommand{\maxd}[0]{\depth_{\text{max}}}

\newcommand{\vocabulary}[0]{\fdset{V}}
\newcommand{\vocabularysize}[0]{|\vocabulary|}

\newcommand{\queries}[0]{\fdset{Q}}

\newcommand{\queriesq}[0]{\queries^{\query}}

\newcommand{\resultsfn}[0]{\fdset{R}}
\newcommand{\results}[1]{\resultsfn_{#1}}

\newcommand{\successorfn}[0]{\fdset{Q}}
\newcommand{\successor}[1]{\successorfn_{#1}}

\newcommand{\query}[0]{q}
\newcommand{\queryidx}[1]{\query_{#1}}
\newcommand{\queryi}[0]{\queryidx{i}}
\newcommand{\queryj}[0]{\queryidx{j}}

\newcommand{\foundquery}[0]{\queryidx{+}}
\newcommand{\firstquery}[0]{\queryidx{0}}

\newcommand{\performance}[0]{\mu}

\newcommand{\qrels}[0]{r}

\newcommand{\numadditions}[0]{n}
\newcommand{\nummerge}[0]{m}
\newcommand{\breadth}[0]{b}
\newcommand{\visited}[0]{\fdset{C}}
\newcommand{\numvisited}[0]{|\visited|}

\begin{document}

\title{Pseudo-Query Reformulation}
\numberofauthors{1}
\author{
\alignauthor Fernando Diaz\\
\affaddr{Microsoft}\\
\email{fdiaz@microsoft.com}
}
\maketitle
\begin{abstract}
	Automatic query reformulation refers to rewriting a user's original query in order to improve the ranking of retrieval results compared to the original query.  We present a general framework for automatic query reformulation based on discrete optimization.  Our approach, referred to as \emph{pseudo-query reformulation},  treats automatic query reformulation as a search problem over the graph of unweighted queries linked by minimal transformations (e.g. term additions, deletions).  This framework allows us to test existing performance prediction methods as heuristics for the graph search process.  We demonstrate the effectiveness of the approach on several publicly available datasets. 
\end{abstract}
\section{Introduction}
\label{sec:introduction}
Most information retrieval systems operate by performing a single retrieval in response to a query. Effective results  sometimes require several manual reformulations by the user \cite{bruza:reformulation,lau:reformulation,huang:reformulation-analysis} or semi-automatic reformulations assisted by the system \cite{huang:query-suggestion,ozertem:ml-query-suggestion,kharitonov:sigir2013}.  Although the reformulation process can be important to the user (e.g. in order to gain perspective about the domain of interest), the process can also lead to frustration and abandonment \cite{feild:frustration}.  

In many ways, the core information retrieval problem is to improve the initial ranking and  user satisfaction and, as a result, reduce the need for reformulations, manual or semi-automatic.  While there have been several advances in learning to rank given a fixed query representation \cite{liu:ltr-book}, there has been somewhat less attention, from a formal modeling perspective, given to automatically reformulating the query before presenting the user with the retrieval results.  One notable exception is pseudo-relevance feedback (PRF), the technique of using terms found in the top retrieved documents to conduct a second retrieval \cite{attar:prf,croft-harper:prf}.  PRF is known to be a very strong baseline.  However, it incurs a very high computational cost because it issues a second, much longer query for retrieval.

\begin{figure}
	\begin{center}
		\includegraphics[width=3in]{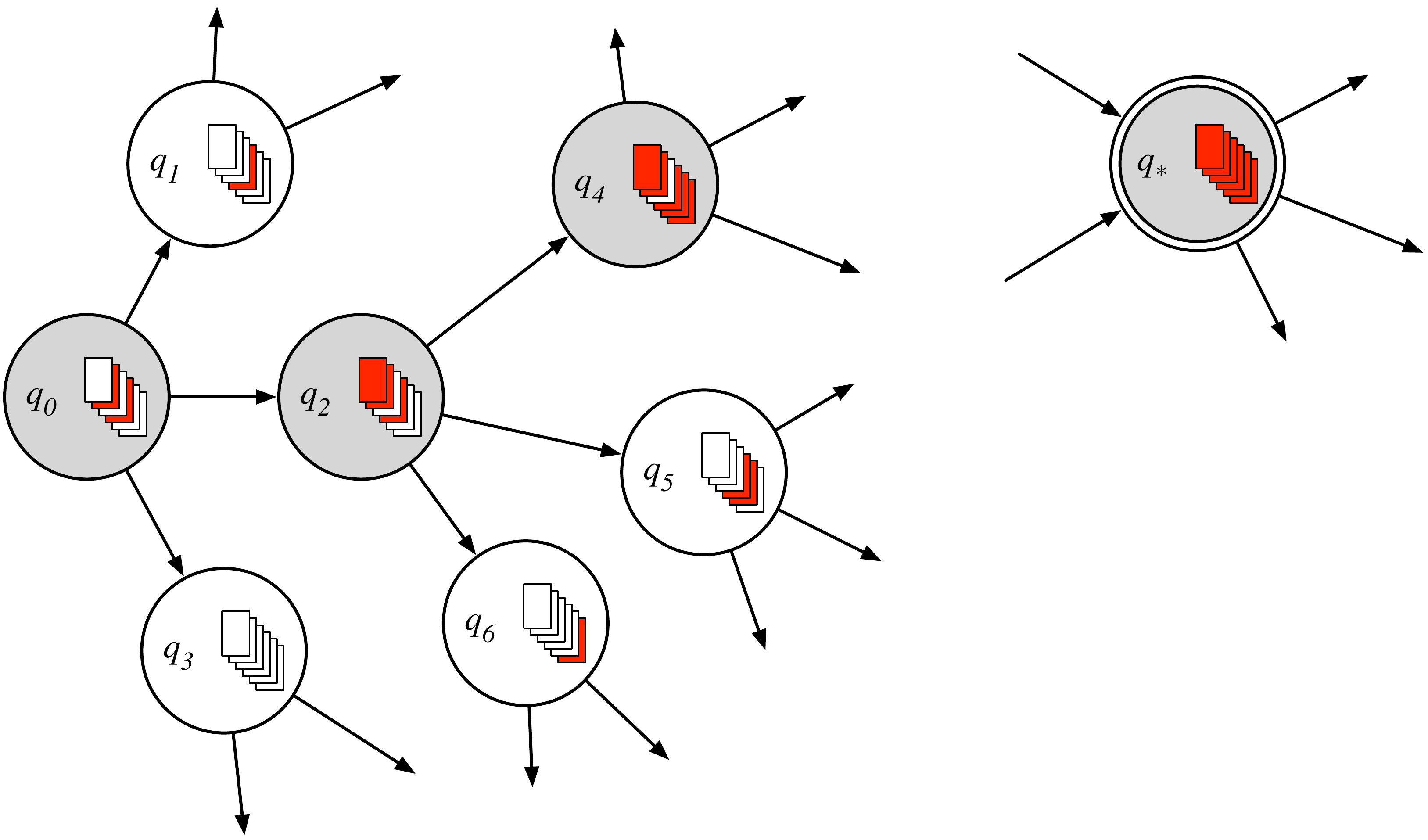}
	\end{center}
	\caption{Query reformulation as graph search.  Nodes represent queries and associated retrieved results.  Relevant documents are highlighted in red.  Edges exist between nodes whose queries are simple reformulations of each other.  The goal of pseudo-query reformulation is to, given a seed query $\firstquery$ by a user, automatically navigate to a better query.}\label{fig:query-flow}
\end{figure}

In this paper, we present an approach to automatic query reformulation which combines the iterated nature of human query reformulation with the automatic behavior of PRF.  We refer to this process as \emph{pseudo-query reformulation} (PQR).  Figure \ref{fig:query-flow} graphically illustrates the intuition behind PQR.  In this figure, each query and its retrieved results are depicted as nodes in a graph.  An edge exists between two nodes, $\queryi$ and $\queryj$, if there is a simple reformulation from $\queryi$ to $\queryj$; for example, a single term addition or deletion.  This simulates the incremental query modifications a user might conduct during a session.  The results in this figure are colored so that red documents reflect relevance.  If we assume that a user is following a good reformulation policy, then, starting at $\firstquery$, she will select reformulations (nodes) which incrementally increase the number of relevant documents.  This is depicted as the path of shaded nodes in our graph.  We conjecture that a user navigates from $\queryi$ to $\queryj$  by using insights from the retrieval results of $\queryi$ (e.g. $\queryj$ includes a highly discriminative term in the results for $\queryi$) or by incorporating some prior knowledge (e.g. $\queryj$ includes a highly discriminative term in general).  PQR is an algorithm which behaves in the same way: issuing a query, observing the results, inspecting possible reformulations, selecting a reformulation likely to be effective, and then iterating.

Several properties make PQR attractive.  First, PQR directly optimizes performance for short, unweighted keyword interaction.  This is important for scenarios where a searcher, human or artificial, is constrained by an API such as those found in many search services provided by general web search engines or social media sites.  This constraint prevents the use of massive query expansion techniques such as PRF. Even if very long queries were supported, most modern systems are optimized (in terms of efficiency and effectiveness) for short queries, hurting the performance of massive query expansion.  Second, our experiments demonstrate that PQR significantly outperforms several baselines, including PRF.  Finally, PQR provides a framework in which to evaluate performance prediction methods in a grounded retrieval task.  

\section{Related Work}
Pseudo-query reformulation draws together three areas of information retrieval: pseudo-relevance feedback, iterative query rewriting, and performance prediction.  Previous research has combined elements of these, but not in the way described in our work.  

Kurland \emph{et al.} present several heuristics for iteratively refining a language model query by navigating document clusters in a retrieval system \cite{kurland:pseudo-query}.  The technique leverages specialized data structures storing document clusters derived from large scale corpus analysis.  While related, the solution proposed by these authors violates assumptions in our problem definition.  First, their solution assumes weighted language model style queries not supported by backends in our scenario.  Second, their solution assumes access to the entire corpus as opposed to a search API.  

Using performance predictors in order to improve ranking has also been studied previously, although in a different context.  Sheldon \emph{et al.} demonstrate how to use performance predictors in order to better merge result lists from pairs of reformulated queries \cite{sheldon:lambdamerge}.  This is, in spirit, quite close to our work and is a special case of PQR which considers only two candidate queries and a single iteration instead of hundreds of candidates over several iterations.  In the context of learning to rank,  performance predictors have been incorporated as ranking signals and been found to be useful \cite{macdonald:usefulness-of-query-features}.  From the perspective of query weighting, Lv and Zhai explored using performance predictors in order to set the optimal interpolation weight in pseudo-relevance feedback \cite{lv:adaptive-prf}.  Similarly Xue and Croft have demonstrated how to use performance predictors in order to improve concept weighting in an inference network model\cite{xue:cikm2010,xue:sigir2012}.  Again, while similar to our work in the use of performance predictors for query reformulation, we focus on the discrete, iterated representation.  The work of Xue and Croft focuses on a single iteration and a weighted representation.  More generally, there has been some interest in detecting the importance of query terms in a long queries or in expanded queries \cite{bendersky:sigir2008,cao:sigir2008,lease:sigir2009,zhao:cikm2010,bendersky:thesis}.

Representing related queries as graphs has been studied extensively.  Early work by Mooers proposed treating the entire space of unweighted queries (i.e. length $|\vocabulary|$ boolean vectors) as a lattice \cite{mooers:lattice-retrieval}.  In the context of web search, Boldi \emph{et al.} studied within-session query reformulations as a graph \cite{boldi:query-flow-graph}.  Other work, such as spreading activation and inference networks as well as term-only graphs are less related although they use a similar formalism.

\section{Motivation}
\label{sec:motivation}

As mentioned earlier, users often reformulate an initial query in response to the system's ranking \cite{bruza:reformulation,lau:reformulation,henzinger:qlog,huang:reformulation-analysis}.  Reformulation actions include adding, deleting, and substituting query words, amongst other transformations.  There is evidence  that manual reformulation can improve the quality of a ranked list for a given information need \cite[Table 5]{huang:reformulation-analysis}.  However, previous research has demonstrated that humans are not as effective as automatic methods in this task \cite{harman:iqe,magennis:IQE,ruthven:term_fb}.  

In order to estimate an upper bound on the potential improvement from reformulation, we propose a simulation of an optimal user's reformulation behavior.  Our simulator is based on query-document relevance judgments, referred to as qrels.  Previous research has used similar techniques to examine the optimality of human reformulation behavior \cite{harman:iqe,magennis:IQE,ruthven:term_fb}.    In this section, we revisit these results with contemporary test collections and retrieval methods.  Unlike this prior work, though, we are not interested in determining the human (in)ability to achieve optimal performance but in gauging the upper bound for PQR.

We sketch our query reformulation simulator in  Figure \ref{fig:user-simulator}.  The simulator is inspired by a model of optimal human search behavior and should not be considered model of any real user.  Our recursive search algorithm uses as input: a reference query $\query$ (e.g. a TREC `title' query), a set of qrels, $\qrels$ for $\query$, a current depth, $\depth$, and a maximum depth, $\maxd$.  The process can be considered a depth-limited graph search by a oracle on the space of queries depicted in Figure \ref{fig:query-flow}.  The simulated search begins by generating a set of  candidate reformulations, $\queriesq$, from an initial query, $\query$.

The next step in our simulation selects the best reformulation from this set of candidates.  We assume that the oracle can measure the performance $\performance$ of the set of candidate reformulations by running each query against the retrieval system and compute a metric such as NDCG with  $\qrels$.  After selecting this query,  we rerun the process on the best reformulation, $\query^*$.  Our search terminates after it reaches a specified depth, $\maxd$. We introduce $\maxd$ in order to limit computation and resource usage.  

Before describing this experiment and results in more detail, we want to make the assumptions of our model clear.  First, the effectiveness of the query found by this simulation is constrained by the query representation.  For example, if our query is an unweighted term vector, then, even if we could exhaustively evaluate all $2^{|\vocabulary|}$ possible queries, we may not find a query achieving the upper bound of the metric (i.e. 1 for most information retrieval metrics).  Therefore, we refer to the \emph{representational upper bound} as the best performance possible using a fixed query representation.  The upper bound found by this simulation is also constrained by the fact that we are performing a local search.  As such, we assume that a better query is reachable from $\firstquery$ through a series of query reformulations.  We do not want to claim that the representational upper bound is reachable or even that a very good query is reachable, only that a better query than $\firstquery$ is reachable.  Fortunately, the previously cited work in human and automatic query reformulation supports this claim.  More subtly, we assume that these `better queries' are reachable through a series of reformulations with increasing performance.  If the better queries are reachable but cannot be navigated to by progressively getting better results, then we will not be able to attain better performance using relevance information.  Unfortunately, this assumption has less justification and we must take it as is.  Note that this assumption does not claim that \emph{all} reformulations $\successor{\firstquery}$ are better than $\firstquery$; only that there exists a better query that is `closer' to even better queries.  Because of these added constraints, we refer to the outcome of this process as the \emph{search-restricted representational upper bound}.

For a random sample of 50 judged training queries, we ran the simulator described in Figure \ref{fig:user-simulator} using the following methods.
The set of candidates consists of all one word deletions and 10 one word additions taken from the 10 ten most frequent words in the retrieval results for $\query$.  
We considered two implementations of $\proc{ScoreQueries}$: oracle prediction and random prediction.  
Oracle prediction computes NDCG@30.  We select this high-precision measure for two reasons.  First, our simulation needs to operate quickly and retrieving shorter lists is much more efficient.  Second,  NDCG is superior at distinguishing high precision runs compared to other measures such as mean average precision \cite{radlinski:metric-sensitivity}.  
Random prediction scores reformulation candidates using a random scalar in the unit range.  
Starting at $\firstquery$, we search up to a depth of four.  Further  details of our corpora and queries can be found in Section \ref{sec:methods}.

\begin{figure}
  \begin{center}
			{\small
	  \begin{codebox}
	        \Procname{$\proc{QRSim}(\query,\depth,\maxd,\qrels)$}

                    \zi $\query$\>\>\Comment current query
    				\zi $\depth$\>\>\Comment current depth
					\zi $\maxd$\>\>\Comment maximum depth
                    \zi $\qrels$\>\>\Comment relevance judgments
					
	        \li \If $\depth=\maxd$
					\li \Then
					\li \Return $\query$
					\End
        \li $\queriesq\gets\proc{GenerateCandidateReformulations}(\query)$
				\li $\performance\gets\proc{ScoreQueries}(\queriesq,\qrels)$
				\li $\query^*\gets\text{argmax}_{\queryi\in\queriesq}\performance_{\queryi}$

				\li \If $(\query^*=\query)$
				\li \Then
        \li \Return $\query$
				\li \Else
        \li \Return $\proc{QRSim}(\query^*,\depth+1,\maxd,\qrels)$

	  \end{codebox}}

\end{center}
\caption{Reformulation simulator.  Given a query $\query$ and query-document relevance judgments $\qrels$, this algorithm will perform gradient ascent on query performance, $\performance$, over the space of query reformulations, $\queries$.  The oracle policy uses $\qrels$ to compute true reformulation performance in $\proc{ScoreQueries}$.  The random policy uses a random number generator for this function.  }\label{fig:user-simulator}
\end{figure}

The results of these experiments (Table \ref{tab:user-simulator-performance}) demonstrate the range of performance for PQR.  Our oracle simulator performs quite well, even given the limited depth of our search.  Performance is substantially better than the baseline, with relative improvements greater than those in published literature.  To some extent this should be expected since the oracle can leverage relevance information.  Surprisingly, though, the algorithm is able to achieve this performance increase by adding and removing a small set of up to four terms.  The poor performance of the random policy suggests that oracle is not just using the terms selected by the initial retrieval to get its boost in performance.
 
\begin{table}
  \begin{center}
  \begin{tabular}{lccc}
		&trec12&robust&web\\
		\hline
QL      &       0.4011        &       0.4260        &       0.1628\\

RM3      &       0.4578        &       0.4312        &       0.1732\\
random       &       0.3162        &       0.2765        &       0.0756\\
$\text{PQR}^*$       &       0.6482        &       0.6214        &       0.3053
       
  \end{tabular}
\end{center}
\caption{NDCG@30 for random (random) and optimal ($\text{PQR}^*$) pseudo-query reformulation compared to query likelihood (QL) and relevance model (RM3).  Datasets are described in Section \ref{sec:data}.  }\label{tab:user-simulator-performance}
\end{table}

Keeping this search-restricted representational upper bound in mind, we would like to develop algorithms that can approximate the behavior of our optimal policy \emph{without having access to any qrels or an oracle}.  The closer our automatic reformulation is to oracle, the better our performance.

\section{Problem Definition}
\label{sec:problemdefinition}

Let $\queries$ be the entire set of queries submittable to a retrieval system.  In the case of unweighted keyword queries, this is all boolean vectors of dimension $\vocabularysize$.  For each query $\query$, we define a set of reformulation candidates, $\successor{\query}$, consisting of all queries reachable by a single term addition or deletion.  For example, the reformulation candidate set for the query {\tt[hello world]} would include {\tt[hello]}, {\tt[world]}, {\tt[hello world \textcolor{BrickRed}{program}]}, {\tt[hello world \textcolor{BrickRed}{song}]}, amongst the $O(\vocabularysize)$ other queries resulting from a single term addition.  Our problem can be stated as follows: given an initial query, $\firstquery$, and access to the candidate generation function, find a query $\foundquery$ that performs better than $\firstquery$.  Performance here is measured by submitting a query to a fixed retrieval system and evaluating results according to a fixed metric (e.g. NDCG@30).  As mentioned earlier, this can be considered a graph search problem where queries are nodes and edges exist between $\query$ and $\successor{\query}$.  Importantly, our algorithm has access to the unweighted keyword retrieval system in order to generate features, but \emph{it never has access to any true relevance information or performance metric}.  Such retrieval services can be found in search APIs such as those provided by major search engines, social media sites, and distributed information retrieval services.

\section{Algorithms}
\label{sec:algorithms}
Conceptually, PQR follows the framework of the simulator from Figure \ref{fig:user-simulator}.  That is, the  algorithm recursively performs candidate generation and candidate scoring within each recursion.  In this section, we will describe candidate set generation (Section \ref{sec:generation}) and candidate scoring (Section \ref{sec:scoring}) along with the graph search algorithm (Section \ref{sec:search}).

\subsection{Generating Candidates}
\label{sec:generation}
Our entire search space can be represented by a very large lattice of queries.  Even if we were performing local graph search, the $O(\vocabularysize)$ edges incident to any one node would make a single iteration computationally intractable.  As a result, we need a method for pruning the full set of reformulation candidates to a smaller set that we can analyze in more detail.  Fortunately, in many cases, we can  establish heuristics so that we only consider those reformulations likely to improve performance.   For example, reformulating the query {\tt [Master theorem]} into {\tt[Master theorem \textcolor{BrickRed}{yak}]} seems unlikely to improve performance if we believe {\tt yak} is unlikely to occur in documents relevant to  {\tt [Master theorem]}.  In our case, given $\query_t$, we consider the following candidates,
\begin{enumerate*}[label=\itshape\alph*\upshape)]
	\item all single term deletions from $\query_t$, and
	\item all single term additions from the $\numadditions$ terms with the highest probability of occurring in relevant documents.
\end{enumerate*}
Since we do not have access to the relevant documents at runtime, we approximate this distribution using the terms occurring in the retrieval for $\query_t$.  Specifically, we select the top $\numadditions$ terms in the relevance model, $\theta_{\results{t}}$, associated with $\query_t$ \cite{victor:RM}.  The relevance model is the  retrieval score-weighted linear interpolation of retrieved document languages models.  We adopt this approach for its computational ease and demonstrated effectiveness in pseudo-relevance feedback.  
\subsection{Scoring Candidates}
\label{sec:scoring}
The candidate generation process described in Section \ref{sec:generation} provides a crude method for pruning the search space.  Based on our observations with the random and oracle policies  in Section \ref{sec:motivation}, we know that inaccurately scoring reformulation candidates can significantly degrade the performance of a scoring algorithm.  In this section, we model the oracle using  established performance prediction signals.  

\subsubsection{Performance Prediction Signals}
\label{sec:scoring:performance-prediction}
Performance prediction refers to the task of ordering a set of queries without relevance information so that the better performing queries are ordered above worse performing queries.  With some exception, the majority of work in this area has focused on ranking queries coming from different information needs (i.e. one query per information need).  We are interested in the slightly different task of ranking many queries for a single information need.  Despite the difference in problem setting, we believe that, with some modifications discussed in Section \ref{sec:scoring:model}, performance predictors can help model the oracle or, more accurately, the true performance of the reformulation.  A complete treatment of related work is beyond the scope of this paper but details of approaches can be found in published surveys (e.g. \cite{hauff:thesis}).  

The set of performance predictors we consider can be broken into three sets: query signals, result set signals, and drift signals.  Throughout this section, we will be describing signals associated with a candidate query $\query$.

\emph{Query signals} refer to properties of the terms in $\query$ alone. These signals are commonly referred to as `pre-retrieval' signals since they can be computed without performing a costly retrieval.  Previous research has demonstrated that queries including non-discriminative terms may retrieve non-relevant results.  The inverse document frequency is one way to measure the discrimination ability of a term and has been used in previous performance prediction work \cite{hauff:pre-retrieval}.  Over all query terms in $\query$, we consider the mean, maximum, and minimum IDF values.  In addition to IDF, we use similarly-motivated signals such as Simplified Clarity (SC) and Query Scope (QS) \cite{he04query-performance-preretrieval-predictors}.

\emph{Result set signals} measure the quality of the documents retrieved by the query.  These signals are commonly referred to as `post-retrieval' signals.  These features include the well-known Query Clarity (QC) measure, defined as the Kullback-Leibler divergence between the language model estimated from the retrieval results, $\theta_{\results{t}}$, and the corpus language model, $\corpuslm$ \cite{clarity}.  In our work, we use $\bhat{\results{t}}{\corpuslm}$, the Bhattacharyya correlation between  the corpus language model and the query language model \cite{Bhattacharyya43}, defined as
\begin{align}
    \bhat{\theta_i}{\theta_j}&=\sum_{w\in\vocabulary}\sqrt{p(w|\theta_i)\times p(w|\theta_j)}
\end{align}
This measure is  in the unit interval and with low values for dissimilar pairs of language models and high values for similar pairs of language models.  The Bhattacharyya correlation has been used effectively other other retrieval tasks \cite{diaz:online-vertical-selection}.  We use the Bhattacharyya correlation  between these two distributions instead of the Kullback-Leibler divergence because the measure is bounded and, as a result, does not need to be rescaled across queries.  We also use the score autocorrelation (SA), a measure of the consistency of scores of semantically related documents \cite{diaz:autocorrelation}.  In our implementation, we again use the Bhattacharyya correlation to measure the similarity between all pairs of documents in $\results{t}$, as represented by their maximum likelihood language models.

\begin{figure}
	\begin{center}

		\begin{subfigure}[b]{1.5in}
			\centering%
			\includegraphics[height=1.25in]{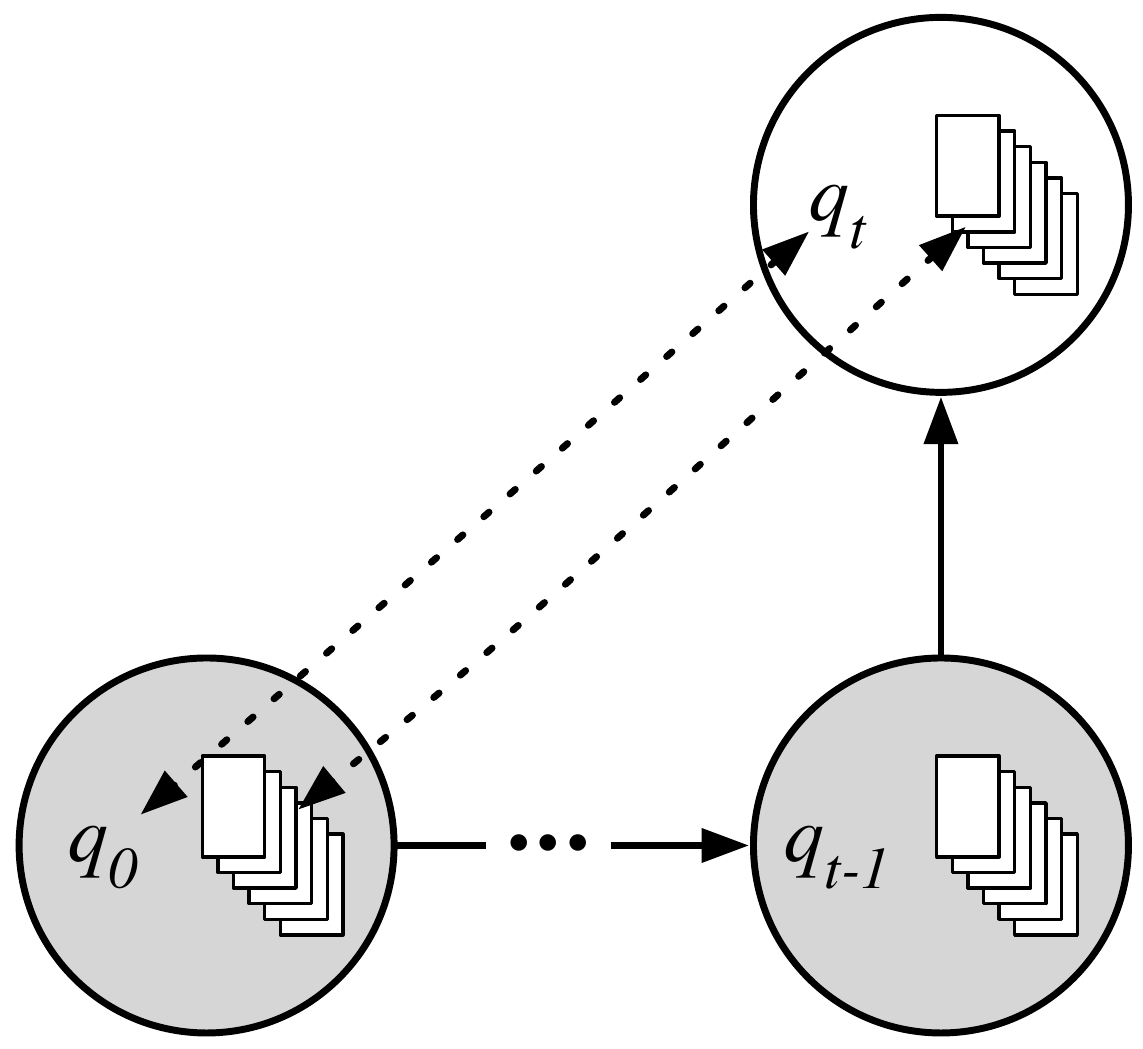}
			\caption{Initial Query}\label{fig:features:q0}
		\end{subfigure}
		\begin{subfigure}[b]{1.5in}
			\centering%
			\includegraphics[height=1.25in]{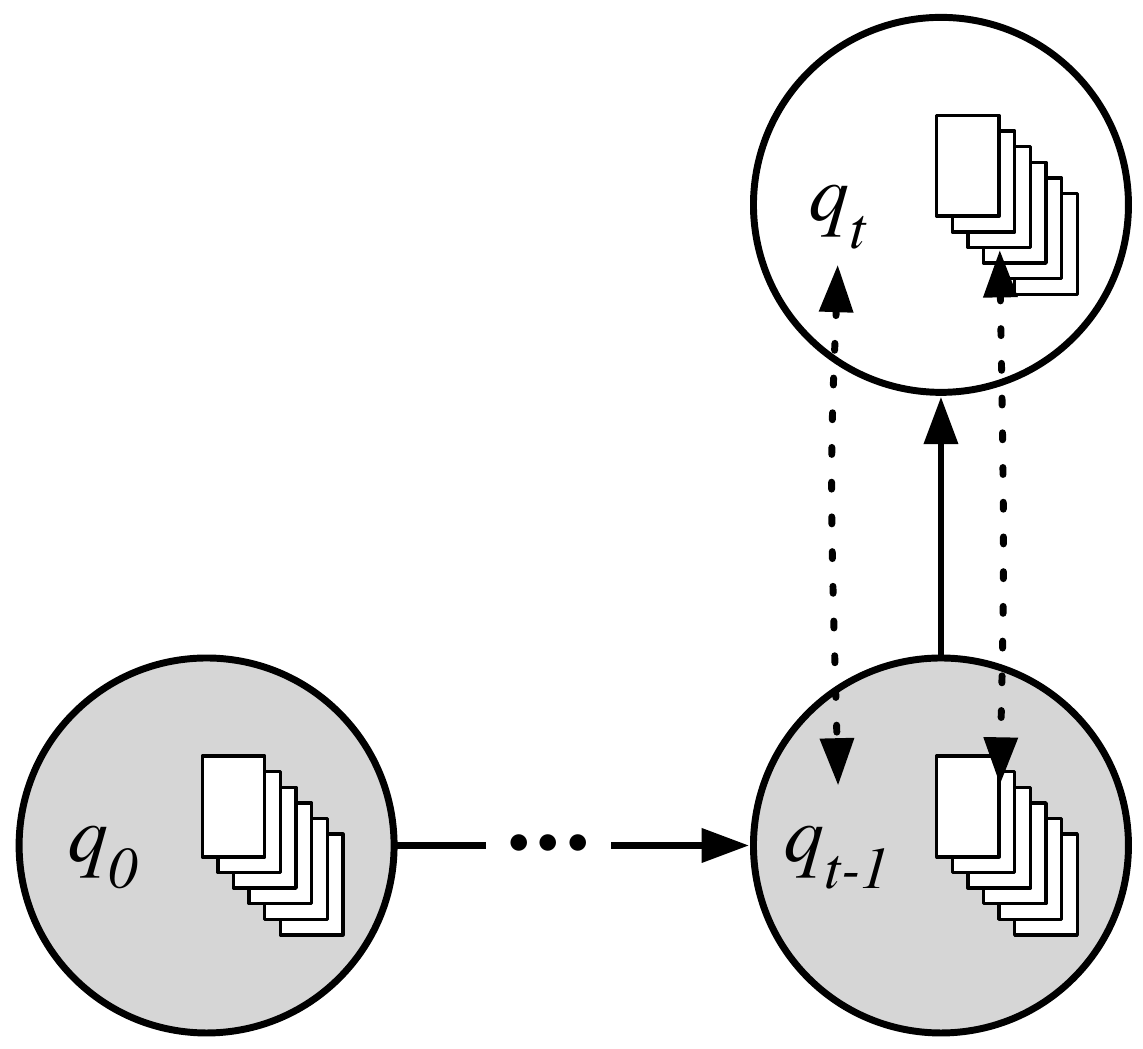}
			\caption{Parent Query}\label{fig:features:qt}
		\end{subfigure}
	\end{center}
	\caption{Drift signal classes.  Signals for $\query_t$ include comparisons with reference queries $\query_{t-1}$ and $\query_0$ to prevent query drift.}\label{fig:features}
\end{figure}

\emph{Drift signals}  compare the current query $\query_{t}$ with its parent $\query_{t-1}$ and the initial query $\firstquery$ (Figure \ref{fig:features}). These signals can serve to anchor our prediction and avoid query drift, situations where a reformulation candidate appears to be high quality but is topically very different from the desired information need.   One way to measure drift is to compute the difference in the query signals for these pairs.  Specifically, we measure the aggregate IDF, SC, and QS values of the deleted, preserved, and introduced keywords.  

We also generate two signals comparing the results sets of these pairs of queries.  The first measures the similarity of the ordering of retrieved documents.  In order to do this, we compute the $\tau$-AP between the rankings \cite{yilmaz:tau-ap}.  The $\tau$-AP computes a position-sensitive version of Kendall's $\tau$ suitable for information retrieval tasks.  The ranking of results for a reformulation candidate with a very high $\tau$-AP will be indistinguishable from those of the reference query; the ranking of results for a reformulation candidate with a very low $\tau$-AP will be quite different from the reference query.    Our second result set signal measures drift by inspecting the result set language models.  Specifically, it computes $\bhat{\theta_{\results{t-1}}}{\theta_{\results{t}}}$, the Bhattacharyya correlation between the result sets.

\subsubsection{Performance Prediction Model}
\label{sec:scoring:model}
With some exception, the majority of performance prediction work has studied predictors independently, without looking at a combinations of signals.  Several approaches to combine predictors focus on regressing against the the absolute performance for a set of training queries \cite{diaz:sigir04,hauf:perfpred-combination}.  This is appropriate when the task is to rank queries from different information needs but it may not be when the task is to predict the performance for reformulation candidates related to the same information need.

\begin{figure}
	\begin{center}
		\begin{subfigure}[b]{3in}
			\centering%
		\includegraphics[width=2.5in]{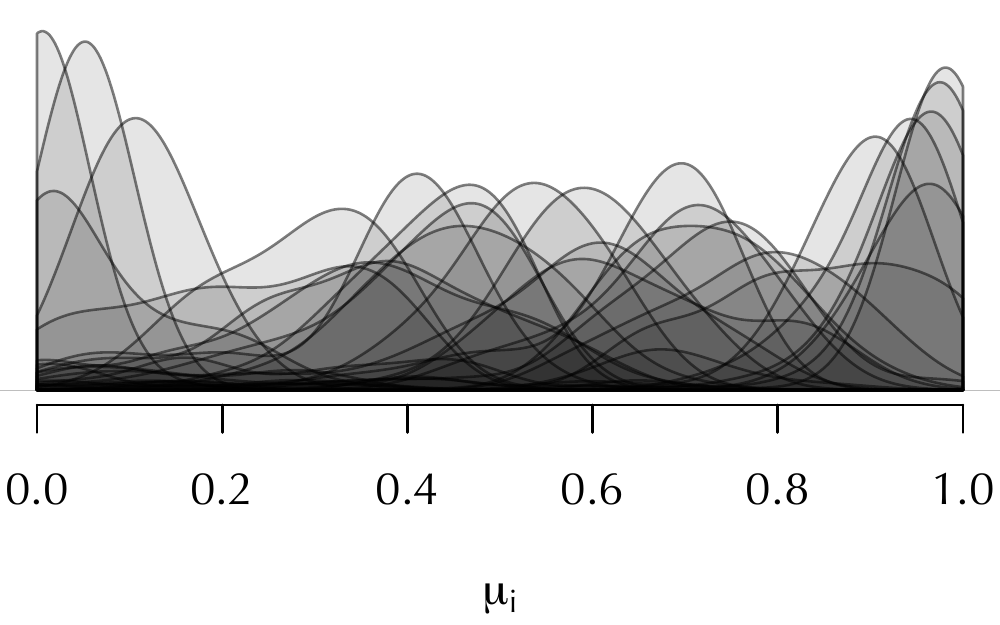}
			\caption{uncentered scores}\label{fig:scores:uncentered}
		\end{subfigure}
		\begin{subfigure}[b]{3in}
			\centering%
		\includegraphics[width=2.5in]{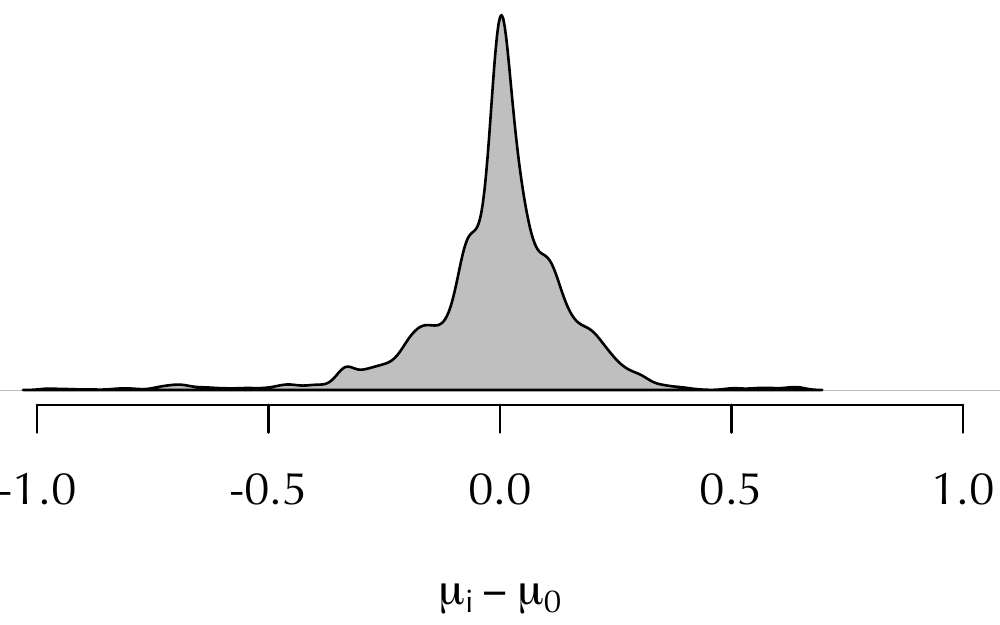}
			\caption{centered scores}\label{fig:scores:centered}
		\end{subfigure}
	\end{center}
	\caption{Distribution of NDCG@30 values for queries visited by the oracle policy for 28 training information needs.  Note that the data for the first plot comes from the oracle policy while the data for the second plot comes from a pseudo-query reformulation policy.}
\end{figure}

In order to demonstrate the problem with regressing against the uncalibrated performance metric for all queries, it is worth inspecting the training data for such an algorithm.  In Figure \ref{fig:scores:uncentered}, we  overlay the distributions of performance metric values for 28 information needs.  Each distribution is a kernel density estimate based on the performance metric values observed when following the graph search algorithm in Section \ref{sec:motivation}.  The figure shows that the relative importance of a reformulation candidate depends strongly on the information need.  Different information needs--as represented by different initial queries--have different mean performance values and, at times, variances.  In fact, the diversity of performance ranges varies dramatically based on the information need, its representation in the corpus, and its complexity;  a good value for one information need may be terrible for another.

Consider the situation where we need to rank a set of reformulation candidates.  The actual value of the metric is less important than the relative value.    One way to address the poorly-calibrated values is to center all performance metric values by subtracting the value of the original query.  The result, a distribution over the relative improvements over $\firstquery$, is presented in Figure \ref{fig:scores:centered}.  This transform is reasonable for our task since it simplifies the regression problem to one of predicting a relative improvement over the baseline as opposed to wasting modeling effort on predicting the absolute performance metric value.  In addition, if the model is accurate, it could provide a convenient method for pruning large areas of the search space predicted to be inferior to $\firstquery$.  

\begin{figure}
	\begin{center}
		\begin{subfigure}[b]{3in}
			\centering%
				\includegraphics[width=2.5in]{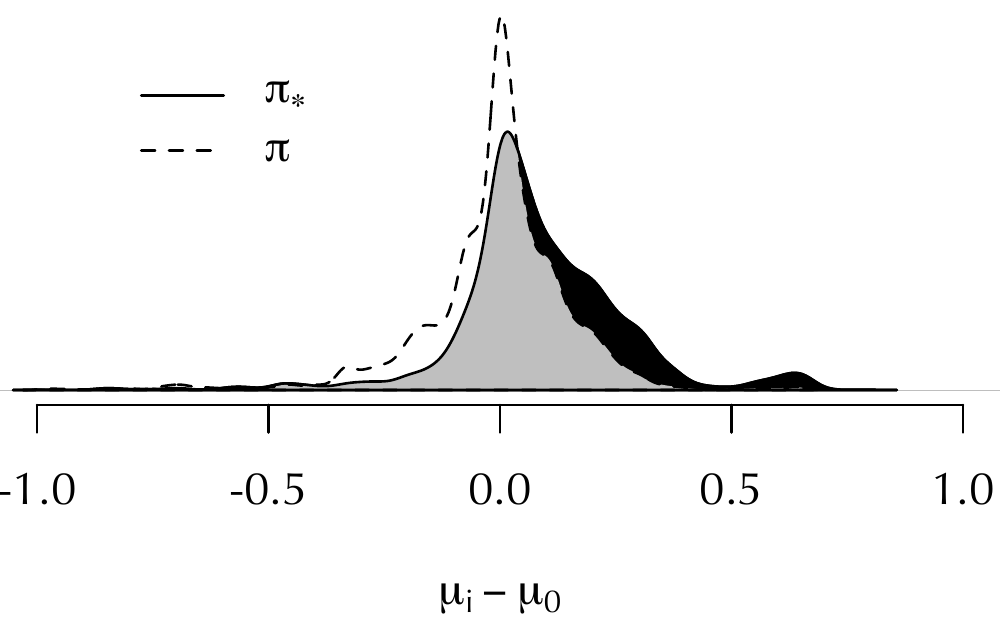}
			\caption{oracle}\label{fig:sampling:oracle}
		\end{subfigure}
		\begin{subfigure}[b]{3in}
			\centering%
				\includegraphics[width=2.5in]{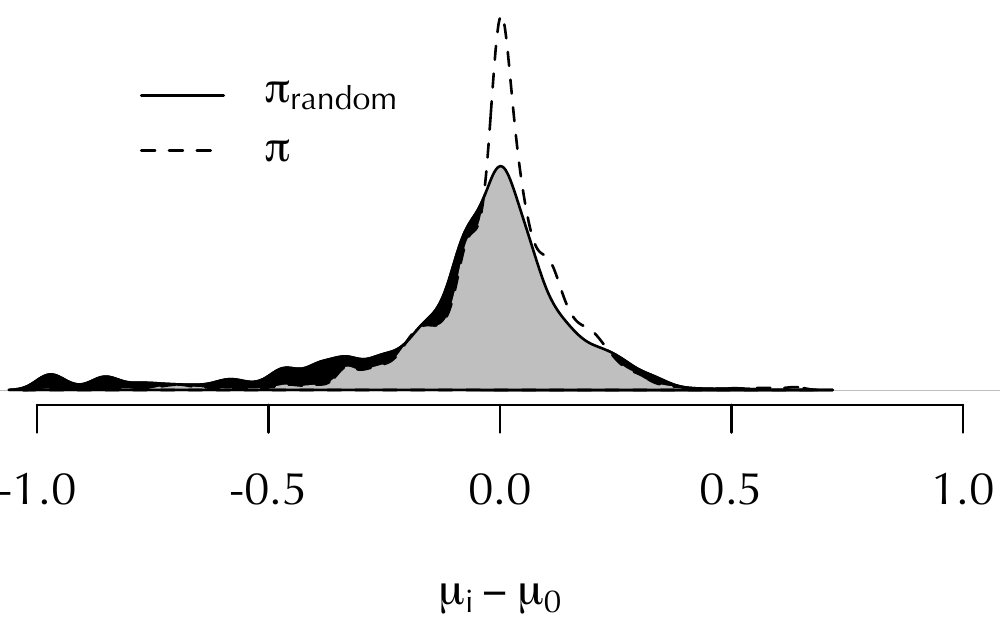}
			\caption{random}\label{fig:sampling:random}
		\end{subfigure}
	\end{center}
	\caption{Score distribution for different data-gathering policies.  The shaded area reflect the distribution with respect to the exploration policy.  The dashed line reflects the distribution with respect to an example solution.  The black area reflects the over-representation by the exploration policy.}\label{fig:sampling}
\end{figure}

Inspecting Figure \ref{fig:scores:centered}, though, also suggests why a regression  against relative performance  which minimizes the mean squared error  may be undesirable.  The distribution is very peaked around the center and a model will be penalized  for poor predictions of reformulation candidates with little or no impact on performance.  In the worst case, the model will predict values close to zero for all reformulation candidates.

Although binning or other techniques can be used to address this situation, we can address this unbalance by simplifying our problem further.  Recall that we really only need a relative ordering of reformulation candidates.  Therefore, we treat this as an ordinal regression problem.  That is, we estimate a model which learns the correct ordering of reformulation candidates for a given information need.  In practice, we train this model using true performance values of candidates encountered throughout a search process started at $\firstquery$; running this process over a number of training $\firstquery$'s results in a large set of training candidates.  Precisely how this training set is collected will be described in the next section.  

Even though we are interested in finding high-performing queries, we will not be biasing our pairwise loss toward the top of the ranked list of candidate queries.  This is because our search algorithm is iterative and observes batches of reformulation candidates at a time, perhaps including highly performing queries, but often not.  We need a model which is accurate for all reformulation candidates, not just the top performing ones.   We are agnostic about the precise functional form of our model and  opt for a linear ranking support vector machine \cite{lee:linear-ranksvm} due to its training and evaluation speed, something we found necessary when conducting experiments at scale.

\begin{figure*}
	\begin{center}
		\begin{subfigure}[b]{3in}
		{\small
  \begin{codebox}
        \Procname{$\proc{QuerySearch}(\query,\depth,\breadth,\maxd,\nummerge)$}
                \zi $\query$\>\>\Comment current query
				\zi $\depth$\>\>\Comment current depth
				\zi $\breadth$\>\>\Comment search breadth
				\zi $\maxd$\>\>\Comment maximum depth
				\zi $\nummerge$\>\>\Comment number of return reformulations
        \li \If $\depth=\maxd$
				\li \Then
				\li \Return $\query$
				\End
        \li $\queriesq\gets\proc{GenerateCandidates}(\query)$
        \li $\fdvec{\tilde{\mu}}\gets\proc{PredictPerformance}(\queriesq)$
        \li $\tilde{\queriesq}\gets\proc{TopQueries}(\queriesq,\fdvec{\tilde{\mu}},\breadth)$
        \li $\hat{\queriesq}\gets\proc{TopQueries}(\queriesq,\fdvec{\tilde{\mu}},\nummerge)$
				\li \For $\queryi \in\tilde{\queriesq}$
				\li \Do
        \li $\hat{\queriesq}\gets\hat{\queriesq}\cup \proc{QuerySearch}(\queryi,\depth+1,\breadth,\maxd,\nummerge)$
				\End
        
        \li $\fdvec{\hat{\mu}}\gets\proc{PredictPerformance}(\hat{\queriesq})$
        \li \Return $\proc{TopQueries}(\hat{\queriesq},\fdvec{\hat{\mu}},\nummerge)$

  \end{codebox}}
	\caption{Query reformulation procedure.}\label{fig:depthfirstsearch}
\end{subfigure}\hspace{.5in}\begin{subfigure}[b]{3in}
	\begin{center}
		\includegraphics[width=1.75in]{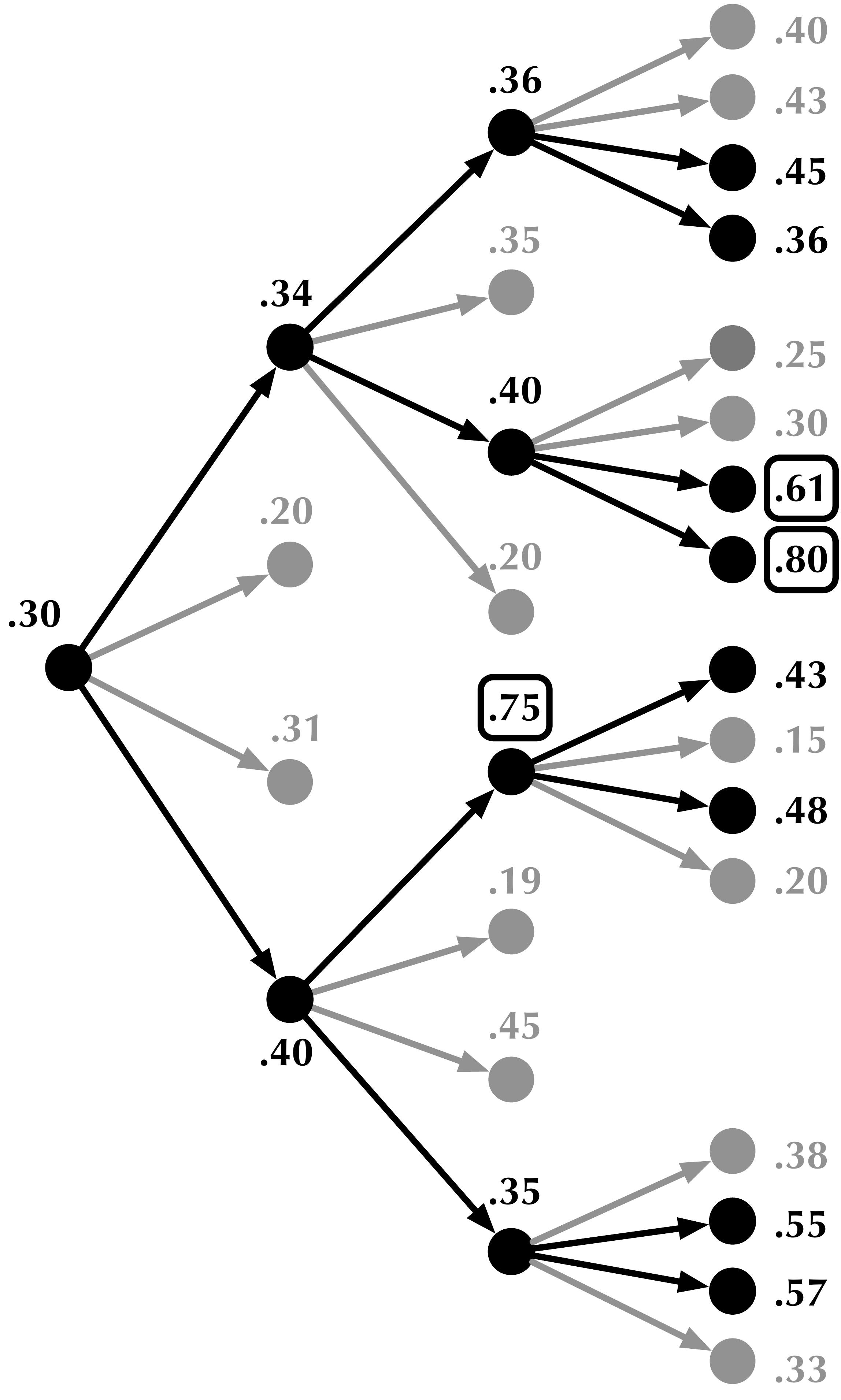}
	\end{center}
	\caption{Illustration of the search process.}\label{fig:scoredNodes}
\end{subfigure}
	\end{center}    
	\caption{The search procedure recursively explores the reformulation graph and returns the top $\nummerge$ highest scoring reformulations inspected. In the illustration, numbers reflect a query's predicted score. The bold nodes represent those nodes selected for expansion.  The highlighted numbers represent the top $\nummerge$ candidates visited throughout the search. }
\end{figure*}
\subsection{Searching Candidates}
\label{sec:search}
Considering the reformulation graph in Figure \ref{fig:query-flow}, the previous two sections explained how to represent the edges (candidate generation) and predict the value of nodes (candidate scoring).  We still need to describe a process for searching for queries starting from $\firstquery$.  We approach this process as a heuristic search problem, using the predicted performance as our heuristic.  Unfortunately, algorithms such as $\text{A}^*$ cannot be reliably used because our heuristic is not admissible.  Similarly, the noise involved in our performance prediction causes greedy algorithms such as beam search or best first search to suffer from local maxima.

Motivated by our search simulator (Figure \ref{fig:user-simulator}), we propose an algorithm that recursively inspects $\numadditions$ reformulation candidates at each $\query_i$ up to a certain depth, $\maxd$.  We present this algorithm in Figure \ref{fig:depthfirstsearch}.  The algorithm differs from our simulation insofar as it executes several reformulation sessions simultaneously, keeping track of those reformulations with the highest predicted effectiveness.  One attractive aspect of our algorithm is the broad coverage of the reformulation space unlikely to be visited in greedier algorithms.

At termination, the algorithm selects a small number ($\nummerge$) of candidate queries visited for final retrieval.  These  $\nummerge$ retrievals are merged using a Borda count algorithm with constituent rankings weighted by predicted performance.  This process allows the algorithm to be more robust to errors in performance prediction.  

The total number of candidates evaluated (Line 4 of Figure \ref{fig:depthfirstsearch}) throughout the search process is approximately,
\begin{align}
    \numvisited&\approx\left\lfloor\frac{\breadth^{\maxd}-1}{\breadth-1}\right\rfloor\numadditions\label{eq:numsearched}
\end{align}
where the approximation error comes from varying initial query length.

\section{Training}
\label{sec:training}
The effectiveness of the search algorithm (Section \ref{sec:search})  critically depends on the reliability of the performance predictor (Section \ref{sec:scoring:model}).  Conversely, the distribution of instances supplied to the performance predictor depends on the decisions made by  the search algorithm.  Therefore, in order to train the performance prediction model, we need to gather example  instances by executing a search and visiting nodes.  Note that, for practical reasons, we cannot possibly gather training signals and targets for \emph{all} queries in our search space.  Even if we could, this set would probably not be representative of the instances the performance prediction model would observe in practice.  For the same reason, we cannot use an arbitrary search policy in order to gather a smaller sample of instances.  To see why this is the case, consider gathering instances for every reformulation candidate inspected by the oracle algorithm described in Section \ref{sec:motivation}.  Even though there will be poorly performing queries in this set of examples, the distribution would  over-represent effective queries because the oracle is guiding the search towards those reformulations.  We demonstrate in Figure \ref{fig:sampling:oracle} where we plot the distribution of centered performance metric values of queries inspected by the oracle compared to a distribution of those inspected by a model used in our experiments.  As expected, the oracle visits a larger number of effective queries on average compared to our example solution.  A model trained on unrepresentative data may be less performant than one trained on data more representative of the queries it will encounter during testing.  At the same time, although sampling with a random policy seems attractive, the distribution of queries inspected here will have the opposite problem.  As shown in Figure \ref{fig:sampling:random}, these queries are will be overrepresent less effective than those visited by the example solution.  

\begin{figure}
	\begin{center}
		\includegraphics[width=2in]{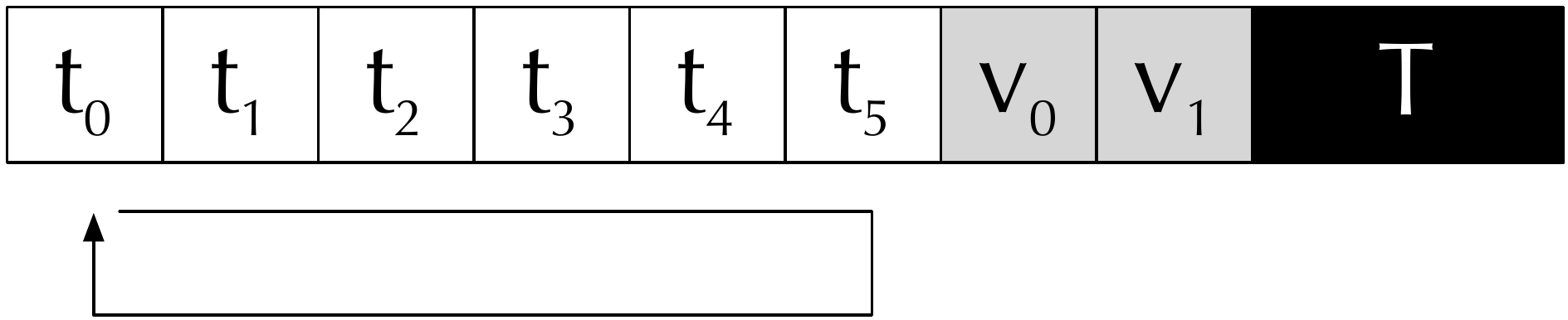}
	\end{center}
	\caption{Partitioning of training ($t_i$), validation ($v_i$), and testing data ($T$).}\label{fig:training-split}
\end{figure}

The solution is to make gather a set of training instances for the performance prediction model which are representative of those  visited by the search at test time.  We accomplish this by gathering training instances using a data-gathering policy that approximates the behavior of our final graph search.  The training operates as follows.  We first partition our training queries into several subsets, $\{t_0,\ldots,t_5\}$; we also partition our validation queries into two subsets $\{v_0,v_1\}$; \emph{our testing queries are left aside for evaluation} (Figure \ref{fig:training-split}).  We then iterate through the training subsets in order.  For each subset, $t_i$, we execute the search algorithm in Figure \ref{fig:depthfirstsearch} using the existing performance prediction model (or the oracle policy if $i=0$).  During the search, we record the feature vector and \emph{true} performance of any encountered query.  This set of $\numvisited\times|t_i|$ instances from $t_i$, can then be used to train a performance prediction model.  The regularization parameter of the SVM is tuned to select the model with the best performance on the validation set, $v_0$.   After this step, we move on to the next training subset, $t_{i+1}$, using newly trained performance prediction model.  As a result of this process, we iteratively accumulate a large set of training instance for the performance predictor representative of instances encountered during the search. Throughout the process we monitor performance on our second validation partition $v_1$.  This method of gathering training representative training data has previously been used in robotics  \cite[Algorithm 3.1]{ross:dagger} and natural language processing \cite[Algorithm 2]{he:dagger}.

We found that making several passes over the training splits improved the model performance on $v_1$.  Therefore, we made several passes over the training splits and selected the model which performed best on $v_1$ for final evaluation.  However, reformulating exactly the same queries in $t_i$ may result in overfitting.  To address this, after the first pass over $t_i$, we deformed the queries using the following procedure.  With equal probability, terms were randomly added or dropped from the original query.  The source of added terms was the true relevance model for the training query.  We applied these perturbations until the Jaccard correlation between the top ten results of the perturbed and unperturbed queries was less than 0.50 and while performance was no less than 75\% of the performance of the unperturbed query.  These conditions ensured that the query was different (in terms of results) but still comparably performant with the unperturbed query.  Similar perturbation processes have been used for computing query-dependent term similarity \cite{kct:perturbation-kernels} and expanding digit recognition data \cite{loosli-canu-bottou-2006}.

\section{Methods}
\label{sec:methods}
\subsection{Data}
\label{sec:data}
We use three standard retrieval corpora for our experiments (Table \ref{tab:corpora}).  Two news corpora, trec12 and robust, consist of large archives of news articles.  The trec12 dataset consists of the Tipster disks 1 and 2 with TREC \emph{ad hoc} topics 51-200.  The robust dataset consists of Tipster disks 4 and 5 with TREC \emph{ad hoc} topics 301-450 and 601-700.  Our web corpus  consists  of the Category B section of the Clue Web 2009 dataset with TREC Web topics 1-200.  We tokenized all corpora on whitespace and then applied Krovetz stemming and removed words  in the SMART stopword list.\footnote{\url{ftp://ftp.cs.cornell.edu/pub/smart/english.stop}}  We further pruned the web corpus of all documents with a Waterloo spam score less than 70.\footnote{\url{https://plg.uwaterloo.ca/~gvcormac/clueweb09spam/}}  We use TREC title queries in all of our experiments.

\begin{table}
	\begin{center}
	\begin{tabular}{lcc}
		&documents&queries\\
		\hline
		trec12&469,949&51-200\\
		robust&528,155&301-450,601-700\\
		web&29,038,227&1-200
	\end{tabular}
	\end{center}
	\caption{Experiment corpora and query sets.  Documents marked as spam removed from web before indexing. }\label{tab:corpora}
\end{table}

We randomly partitioned the queries into three sets: 60\% for training, 20\% for validation, and 20\% for testing.  We repeated this random split procedure five times and present results averaged across the test set queries.

\subsection{Implementation}
\label{sec:implementation}
All indexing and retrieval was conducted using indri 5.7.\footnote{\url{http://www.lemurproject.org/indri/}}  Our SVM models were trained using liblinear 1.95.\footnote{\url{http://www.csie.ntu.edu.tw/~cjlin/libsvmtools/#large_scale_ranksvm}}  We evaluated final retrievals using NIST trec\_eval 9.0.\footnote{\url{http://trec.nist.gov/trec_eval/}}  In order to support large parameter sweeps, each query reformulation in PQR performed a \emph{re-ranking} of the documents retrieved by $\firstquery$ instead of a \emph{re-retrieval} from the full index.  Pilot experiments found that the effectiveness of re-retrieval was comparable with that of re-ranking though re-retrieval incurred  \emph{much} higher latency.  

\subsection{Parameters}
\label{sec:tuning}
Aside from the performance prediction model, our algorithm has the following free parameters: the number of term-addition candidates per query ($\numadditions$), the number of candidates to selection per query ($\breadth$), and the maximum search depth ($\maxd$).  Combined,  the automatic reformulation and the multi-pass training resulted in computationally expensive processes whose runtime is sensitive to these parameters.  Consequently, we fixed our parameter settings at relatively modest numbers ($\numadditions=10,\breadth=3,\maxd=4$) and leave a more thorough analysis of sensitivity for an extended manuscript.  Although these numbers may seem small, we remind the reader that this results in roughly $\numvisited\approx 800$ reformulations considered within the graph search for a single $\firstquery$ (Equation \ref{eq:numsearched}).  The number of candidates to merge ($\nummerge$) is tuned throughout training on the validation set $v_0$ and ranges from five to twenty.

The query likelihood baseline used Dirichlet smoothing with parameter tuned on the full training set using a range of values from 500 through 5000.  The parameters of the relevance model baseline (RM3) were also tuned on the full training set.  The range of feedback terms considered was $\{5,10,25,50,75,100\}$; the range of feedback documents was $\{5,25,50,75,100\}$; the range of $\lambda$ was $[0,1]$ with a step size of $0.1$.  

All  runs, including baselines, optimized NDCG@30.

\section{Results}
\label{sec:results}
\begin{table*}
	 \caption{Comparison of PQR to query likelihood (QL) and relevance model (RM3) baselines for our datasets.  Statistically significant difference with respect to QL ($\blacksquare$: better; $\square$: worse) and RM3 ($\blacklozenge$: better; $\lozenge$: worse) using a Student's paired $t$-test  ($p<0.05$ with a Bonferroni correction).  The best performing run is presented in bold.  All runs have parameters tuned for NDCG@30 on the validation set.}\label{tab:results}
	\begin{center}
\begin{tabular}{lcccccc}
   &     NDCG@5  &   NDCG@10  &   NDCG@20  &   NDCG@30  &   NDCG  &   MAP \\
   \hline
   trec12\\
	QL	&	$0.5442^{\phantom{\square}\phantom{\lozenge}}$	&	$0.5278^{\phantom{\square}\phantom{\lozenge}}$	&	$0.5066^{\phantom{\square}\phantom{\lozenge}}$	&	$0.4835^{\phantom{\square}\phantom{\lozenge}}$	&	$0.5024^{\phantom{\square}\phantom{\lozenge}}$	&	$0.2442^{\phantom{\square}\phantom{\lozenge}}$	\\
	RM3	&	$\mathbf{0.6465}^{\blacksquare\phantom{\lozenge}}$	&	$\mathbf{0.6113}^{\blacksquare\phantom{\lozenge}}$	&	$\mathbf{0.5796}^{\blacksquare\phantom{\lozenge}}$	&	$\mathbf{0.5627}^{\blacksquare\phantom{\lozenge}}$	&	$\mathbf{0.5300}^{\blacksquare\phantom{\lozenge}}$	&	$\mathbf{0.2983}^{\blacksquare\phantom{\lozenge}}$	\\
	random	&	$0.5690^{\phantom{\square}\lozenge}$	&	$0.5563^{\phantom{\square}\lozenge}$	&	$0.5257^{\phantom{\square}\lozenge}$	&	$0.5089^{\phantom{\square}\lozenge}$	&	$0.5120^{\blacksquare\lozenge}$	&	$0.2653^{\blacksquare\lozenge}$	\\
	PQR	&	$0.6112^{\blacksquare\lozenge}$	&	$0.5907^{\blacksquare\phantom{\lozenge}}$	&	$0.5630^{\blacksquare\phantom{\lozenge}}$	&	$0.5419^{\blacksquare\lozenge}$	&	$0.5216^{\blacksquare\lozenge}$	&	$0.2819^{\blacksquare\lozenge}$	\\
\\
robust\\
	QL	&	$0.4874^{\phantom{\square}\phantom{\lozenge}}$	&	$0.4559^{\phantom{\square}\phantom{\lozenge}}$	&	$0.4306^{\phantom{\square}\phantom{\lozenge}}$	&	$0.4172^{\phantom{\square}\phantom{\lozenge}}$	&	$0.5419^{\phantom{\square}\phantom{\lozenge}}$	&	$0.2535^{\phantom{\square}\phantom{\lozenge}}$	\\
	RM3	&	$0.4888^{\phantom{\square}\phantom{\lozenge}}$	&	$0.4553^{\phantom{\square}\phantom{\lozenge}}$	&	$0.4284^{\phantom{\square}\phantom{\lozenge}}$	&	$0.4176^{\phantom{\square}\phantom{\lozenge}}$	&	$0.5462^{\phantom{\square}\phantom{\lozenge}}$	&	$0.2726^{\blacksquare\phantom{\lozenge}}$	\\
	random	&	$0.4240^{\square\lozenge}$	&	$0.3967^{\square\lozenge}$	&	$0.3675^{\square\lozenge}$	&	$0.3588^{\square\lozenge}$	&	$0.5143^{\square\lozenge}$	&	$0.2352^{\square\lozenge}$	\\
	PQR	&	$\mathbf{0.5009}^{\phantom{\square}\phantom{\lozenge}}$	&	$\mathbf{0.4713}^{\blacksquare\blacklozenge}$	&	$\mathbf{0.4438}^{\blacksquare\blacklozenge}$	&	$\mathbf{0.4315}^{\blacksquare\blacklozenge}$	&	$\mathbf{0.5498}^{\blacksquare\phantom{\lozenge}}$	&	$\mathbf{0.2736}^{\blacksquare\phantom{\lozenge}}$	\\
\\
web\\
	QL	&	$0.2206^{\phantom{\square}\phantom{\lozenge}}$	&	$0.2250^{\phantom{\square}\phantom{\lozenge}}$	&	$0.2293^{\phantom{\square}\phantom{\lozenge}}$	&	$0.2315^{\phantom{\square}\phantom{\lozenge}}$	&	$0.3261^{\phantom{\square}\phantom{\lozenge}}$	&	$0.1675^{\phantom{\square}\phantom{\lozenge}}$	\\
	RM3	&	$0.2263^{\phantom{\square}\phantom{\lozenge}}$	&	$0.2273^{\phantom{\square}\phantom{\lozenge}}$	&	$0.2274^{\phantom{\square}\phantom{\lozenge}}$	&	$0.2316^{\phantom{\square}\phantom{\lozenge}}$	&	$\mathbf{0.3300}^{\blacksquare\phantom{\lozenge}}$	&	$\mathbf{0.1736}^{\blacksquare\phantom{\lozenge}}$	\\
	random	&	$0.1559^{\square\lozenge}$	&	$0.1562^{\square\lozenge}$	&	$0.1549^{\square\lozenge}$	&	$0.1537^{\square\lozenge}$	&	$0.2790^{\square\lozenge}$	&	$0.1157^{\square\lozenge}$	\\
	PQR	&	$\mathbf{0.2528}^{\blacksquare\blacklozenge}$	&	$\mathbf{0.2501}^{\blacksquare\blacklozenge}$	&	$\mathbf{0.2493}^{\blacksquare\blacklozenge}$	&	$\mathbf{0.2435}^{\blacksquare\phantom{\lozenge}}$	&	$\mathbf{0.3300}^{\phantom{\square}\phantom{\lozenge}}$	&	$0.1690^{\phantom{\square}\phantom{\lozenge}}$	\\
\end{tabular}

	 \end{center}
\end{table*}

We present the results for our  experiments in Table \ref{tab:results}.  Our first baseline, query likelihood (QL) reflects the performance of $\firstquery$ alone and represents an algorithm which is representationally comparable with PQR insofar as it also retrieves using a short, unweighted query.  Our second baseline, the relevance model (RM3) reflects the performance of a strong algorithm that also uses the retrieval results to improve performance, although with much richer representational power (the optimal number of terms often hover near 75-100).  As expected, RM3 consistently outperforms QL in terms of MAP.  And while the performance is superior across all metrics for trec12, RM3 is statistically indistinguishable from QL for higher precision metrics on our other two data sets.  The random policy, which replaces our performance predictor with random scores, consistently underperforms both baselines for robust and web.  Interestingly, this algorithm is statistically indistinguishable from QL for trec12, suggesting that this corpus may be easier than  others.  

Next, we turn to the performance of PQR.  Across all corpora and across almost all metrics, PQR significantly outperforms QL.  While this baseline might be considered low, it is a representationally fair comparison with PQR.  So, this result demonstrates the ability of PQR to find more effective reformulations than $\firstquery$.  The underperformance of the random algorithm signifies that the effectiveness of PQR is attributable to the performance prediction model as opposed to a merely walking on the reformulation graph.  That said, PQR is statistically indistinguishable from QL for higher recall metrics on the web corpus (NDCG and MAP).  In all likelihood, this results from the optimization of NDCG@30, as opposed to higher recall metrics.  This outcome is amplified when we compare PQR to RM3.  For the robust and web datasets, we notice PQR significantly outperforming RM3 for high precision metrics but showing weaker performance for high recall metrics.  We point out that PQR performs weaker than RM3 for trec12.  This might be explained by the easier nature of the corpus combined with the richer representation of the RM3 model.

We can inspect the coefficient values to determine the importance of individual signals in performance prediction.  In Table \ref{tab:feature-weights}, we present the most important signals for each of our experiments.  Because our results are averaged over several runs, we selected the signals most often occurring amongst the highest weighted in these runs, using the final selected model (see Section \ref{sec:training}).  Interestingly, many of the top ranked signals are our drift features which compare the language models and rankings of the candidate result set with those of its parent and the first query.  This suggests that the algorithm is successfully preventing query drift by promoting candidates that retrieve results similar to the original and parent queries.   On the other hand, the high weight for Clarity suggests that PQR is simultaneously balancing ranked list refinement with ranked list anchoring.  

\begin{table}
	 \caption{Top five highest weighted signals for each experiment.  For each run in each experiment, we ranked all signals by the magnitude of their associated weight in the linear model.  We aggregated these rankings and present the signals ranked by frequency in the top five signals across runs. }\label{tab:feature-weights}
	\begin{center}
	\begin{tabular}{ccc}
		trec12&robust&web\\
		\hline
$\bhat{\theta_{\results{0}}}{\theta_{\results{t}}}$ & $\bhat{\theta_{\results{0}}}{\theta_{\results{t}}}$ & $\tau_{\text{AP}}(\results{0},\results{t})$ \\
$\bhat{\theta_{\results{t-1}}}{\theta_{\results{t}}}$ & $\bhat{\theta_{\results{t-1}}}{\theta_{\results{t}}}$ & $\bhat{\theta_{\results{0}}}{\theta_{\results{t}}}$ \\
$\tau_{\text{AP}}(\results{0},\results{t})$ & Clarity & $\tau_{\text{AP}}(\results{t-1},\results{t})$ \\
$\tau_{\text{AP}}(\results{t-1},\results{t})$ & $\tau_{\text{AP}}(\results{t-1},\results{t})$ & $\bhat{\theta_{\results{t-1}}}{\theta_{\results{t}}}$ \\
Clarity & maxIDF & Clarity
	 \end{tabular}
	 \end{center}
\end{table}

\section{Discussion}
\label{sec:discussion}
Although QL is the appropriate baseline for PQR, comparing PQR performance to that of RM3 helps us understand where improvements may be originating.  The effectiveness of RM3 on trec12 is extremely strong, demonstrating statistically superior performance to PQR on many metrics.  At the same time, the absolute metrics for QL on these runs is also higher than on the other two collections.  This suggests that part of the effectiveness of RM3 results from the strong initial retrieval (i.e. QL).  As mentioned earlier, the strength of the random run separately provides evidence of the initial retrieval's strength.  Now, if the initial retrieval uncovered significantly more relevant documents, then RM3 will estimate a language model very close to the true relevance model, boosting performance.  Since RM3 allows a long, rich, weighted query, it follows that it would outperform PQR's constrained representation.  That said, it is remarkable that PQR achieves comparable performance to RM3 on many metrics with at most $|\firstquery|+\maxd$ words.

The weaker performance for high-recall metrics was somewhat disappointing but should be expected given our optimization target (NDCG@30).  Post-hoc experiments demonstrated that optimizing for MAP boosted the performance of PQR to 0.1728 on web, resulting in statistically indistinguishable performance with RM3.  Nevertheless, we are not certain that human query reformulation of the type encountered in general web search would improve high recall metrics since users in that context rarely inspect deep into the ranked list.

One of the biggest concerns with PQR is efficiency.  Whereas our QL baseline ran in a 100-200 milliseconds, PQR ran in 10-20 \emph{seconds}, even using the re-ranking approach.  However, because of this approach, our post-retrieval costs scale modestly as corpus size grows, especially compared to massive query expansion techniques like RM3.  To understand this observation, note that issuing a long RM3 query results in a huge slowdown in performance due to the number of postings lists that need to be evaluated and merged.  We found that for the web collection, RM3 performed quite slow, often taking \emph{minutes} to complete long queries.  PQR, on the other hand, has the same overhead as RM3 in terms of an initial retrieval and fetching document vectors.  After this step, though, PQR only needs to access the index for term statistic information, not a re-retrieval.  Though even with our speedup, PQR is unlikely to be helpful for realtime, low-latency retrieval.  However, there are several situations where such a technique may be permissible.  For example, `slow search' refers to search situations where users tolerate latency in order to receive better results \cite{teevan:slow-search}.  Another situation is  document filtering, where the user has a standing query for a certain topic and the system can optimize its query representation during indexing lulls.  More generally, this technique is also valuable for any distributed information retrieval problem with APIs constrained to unweighted queries.

\section{Conclusion}
\label{sec:conclusion}
The positive results on three separate corpora provide evidence that PQR is a framework worth investigating further.  In terms of candidate generation, we considered only very simple word additions and deletions while previous research has demonstrated the effectiveness of applying multiword units (e.g. ordered and unordered windows) \cite{metzler:lce}.  Beyond this, we can imagine applying more sophisticated operations such as filters, site restrictions, or time ranges.  While it would increase our query space, it may also allow for more precise and higher precision reformulations.  In terms of candidate scoring, we found that our novel drift signals allowed for effective query expansion.  We believe that PQR provides a framework for developing other performance predictors in a grounded retrieval model.  In terms of graph search, we believe that other search strategies might result in more effective coverage of the space.  

We would like to return to our original motivation: mimicking human reformulation.  We have developed framework for learning reformulation behavior from an oracle.  In many situations, as with production web search engines, we have access to human reformulation behavior.  Given such data, we could train a PQR model directly on human behavior.  Although prior work demonstrates the poor performance of human reformulation, we are interested in exploring the effects on our trained models.

\bibliographystyle{abbrv}

\balancecolumns
\end{document}